\documentclass[aps,prb,twocolumn,superscriptaddress,amsmath]{revtex4-1}
\usepackage{overpic} 
\usepackage{braket}
\usepackage{graphicx}
\usepackage{xcolor}
\usepackage{pstricks}
\usepackage{subfig}
\usepackage{bbold}
\usepackage{mathbbol}
\usepackage{bbm}
\usepackage{grffile}
\begin{document}

%\preprint{}

%\preprint{APS/123-QED}

\title{Charge Transport in C$_{60}$-based Single-Molecule Junctions with Graphene Electrodes}



%\thanks{}

\author{S.\ Leitherer}
\affiliation{
Institute for Theoretical Physics and Interdisciplinary Center for Molecular Materials, \\
Friedrich-Alexander-University Erlangen-N\"urnberg (FAU),\\
Staudtstr.\ 7/B2, D-91058 Erlangen, Germany
}
\author{P.\ B.\ Coto}
\affiliation{
Institute for Theoretical Physics and Interdisciplinary Center for Molecular Materials, \\
Friedrich-Alexander-University Erlangen-N\"urnberg (FAU),\\
Staudtstr.\ 7/B2, D-91058 Erlangen, Germany
}
\author{K.\ Ullmann}
\affiliation{
Chair of Applied Physics and Interdisciplinary Center for Molecular Materials, \\
Friedrich-Alexander-University Erlangen-N\"urnberg (FAU),\\
Staudtstr.\ 7/A3, D-91058 Erlangen, Germany
}
\author{H.\ B.\ Weber}
\affiliation{
Chair of Applied Physics and Interdisciplinary Center for Molecular Materials, \\
Friedrich-Alexander-University Erlangen-N\"urnberg (FAU),\\
Staudtstr.\ 7/A3, D-91058 Erlangen, Germany
}
\author{M.\ Thoss}
\affiliation{
Institute for Theoretical Physics and Interdisciplinary Center for Molecular Materials, \\
Friedrich-Alexander-University Erlangen-N\"urnberg (FAU),\\
Staudtstr.\ 7/B2, D-91058 Erlangen, Germany
}

\date{\today}

\begin{abstract}
We investigate charge transport in C$_{60}$-based single-molecule junctions with graphene electrodes  employing a combination of density functional theory (DFT) electronic structure calculations and 
Landauer transport theory. In particular, the dependence of the transport properties on the conformation of the molecular bridge and the type of termination of the graphene electrodes is investigated. Furthermore, electron pathways through the junctions are analyzed using the theory of local currents. The results reveal, in agreement with previous experiments, a pronounced dependence of the transport properties on the bias polarity, which is rationalized in terms of the electronic structure of the molecule. 
It is also shown that the edge states of zigzag-terminated graphene induce additional transport channels, which dominate transport at small voltages. The importance of the edge states for transport depends profoundly on the interface geometry of the junctions.
\end{abstract}

%\keywords{}

\maketitle


%%%MAIN TEXT%%%%

\section{Introduction}

The two-dimensional carbon material graphene is characterized by a number of outstanding properties including high electron mobility, high thermal conductivity, and mechanical stability, which make it a promising material for electronic devices.\cite{Geim07} An important example is the use of graphene as alternative material for electrodes in nanoelectronic applications, replacing conventional metal leads.\cite{Geim09}
This application appears particularly promising for single-molecule junctions in the field of molecular electronics, where the replacement of metal electrodes by graphene could help to overcome several limitations. Due to its two-dimensional geometry, its small optical absorption cross section and high thermal conductivity, graphene electrodes can facilitate the access to the junction using light fields or scanning probe tips, which will provide a wealth of experimental control parameters. Furthermore, graphene edges 
may reduce the uncertainty about the molecule-electrode
contact geometry, a major drawback of metal electrodes.\cite{Petervalvi}

\begin{figure}[b]
 	\includegraphics[width=0.9\linewidth]{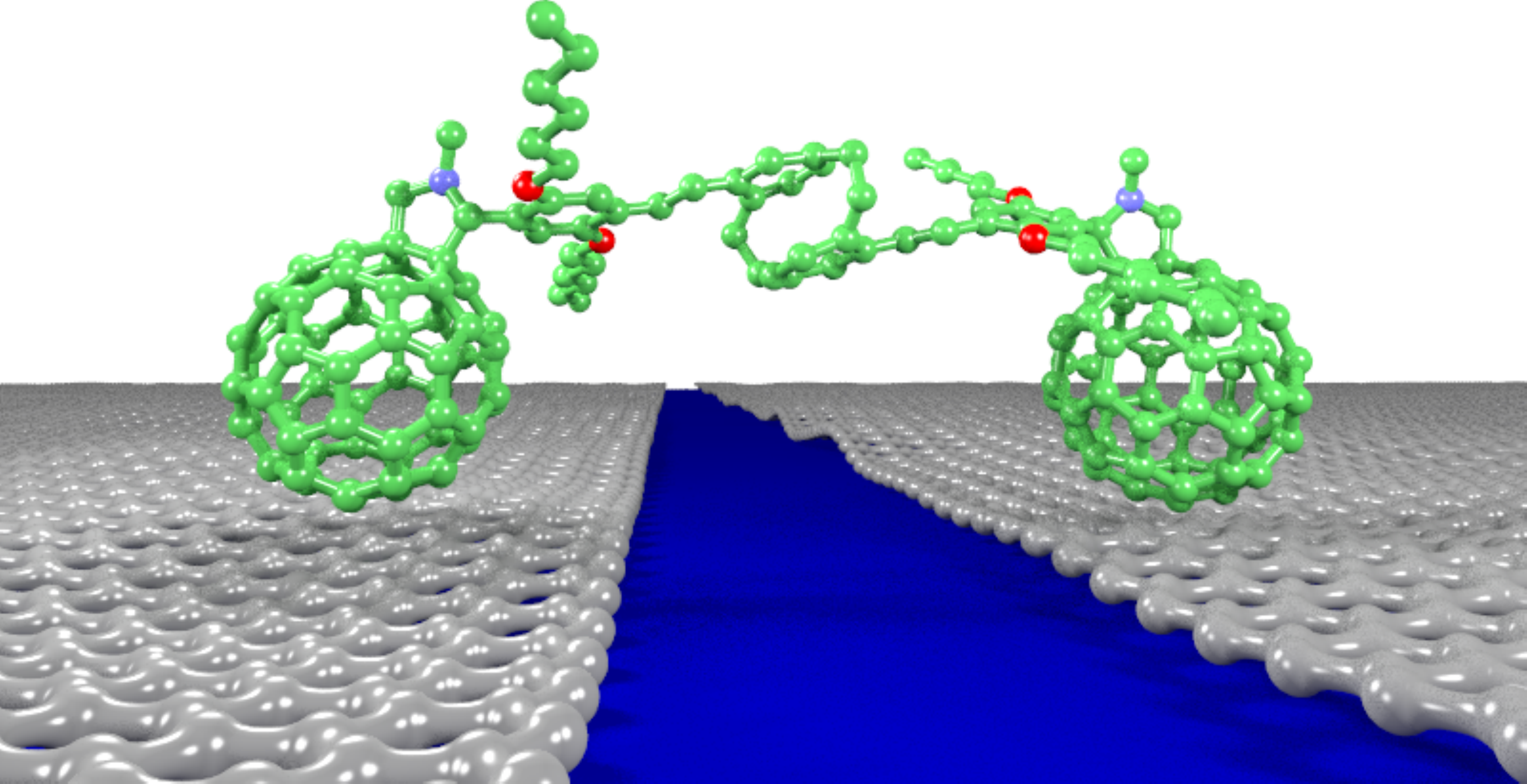}\vspace{0.5cm}
 	\caption{C$_{\texttt{60}}$-endcapped molecule-graphene junction.}
 	\label{fig:001}
\end{figure}

Transport through graphene-based nanojunctions has been the focus of several theoretical investigations, \cite{tan2011,Nikolic2012,Gunst2016} 
where phenomena such as spin transport, heat transport or the role of phonons were studied.
The theoretical studies have also shown that it is crucial for the development of graphene-based nanoelectronics to control the edge structure.\cite{engelund2010,Jia2009}
Specifically, the role of zigzag edge states of graphene has been examined in detail. \cite{Ivan2013,Motta,Ryndyk2012} 

Recent experiments have demonstrated the use of graphene electrodes in single molecule junctions.\cite{Ferry11,Ullmann15,Mol15,Burzuri16,Jia2016}
In Ref.\ \citenum{Ullmann15}, Ullmann and co-workers reported on experiments using epitaxial graphene electrodes and a C$_{\texttt{60}}$-endcapped molecular bridge. The results of that study showed a striking similarity of the transport properties using graphene and gold electrodes, demonstrating that transport in this system is dominated by the molecular bridge itself. Furthermore strongly asymmetric current-voltage characteristics and switching behavior were found, which were associated to the existence of different conformers.

In this paper, we provide a detailed theoretical analysis of transport in this class of C$_{\texttt{60}}$-endcapped molecule-graphene nanojunctions. The study employs a combination of density functional theory (DFT) and Landauer transport theory. In addition, the theory of local currents\cite{solomon08,leitherer14} is used to identify pathways of electrons through the molecular junctions. 

The article is organized as follows: In Sec.\ 
\ref{sec:methods_setup}, we introduce  the class of junctions investigated throughout the paper. The 
theoretical methodology  is 
outlined in Secs.\ \ref{sec:methods_elstruc} - \ref{sec:methods_loccurr},
including electronic structure calculations, the Landauer transport approach as well as the theory of local currents.
Sec.\ \ref{sec:results} presents the results of the simulations obtained
for various types of junctions, which differ in the conformation of the molecule and the termination of the graphene electrodes. 
We discuss the transport properties
and their relation to the electronic states of the molecule as well as the influence of the interface geometry and the role of graphene edge states.
Additionally, local pathways of electrons in the junction are analyzed.

\begin{figure*}
\centering
  \includegraphics[scale=0.27]{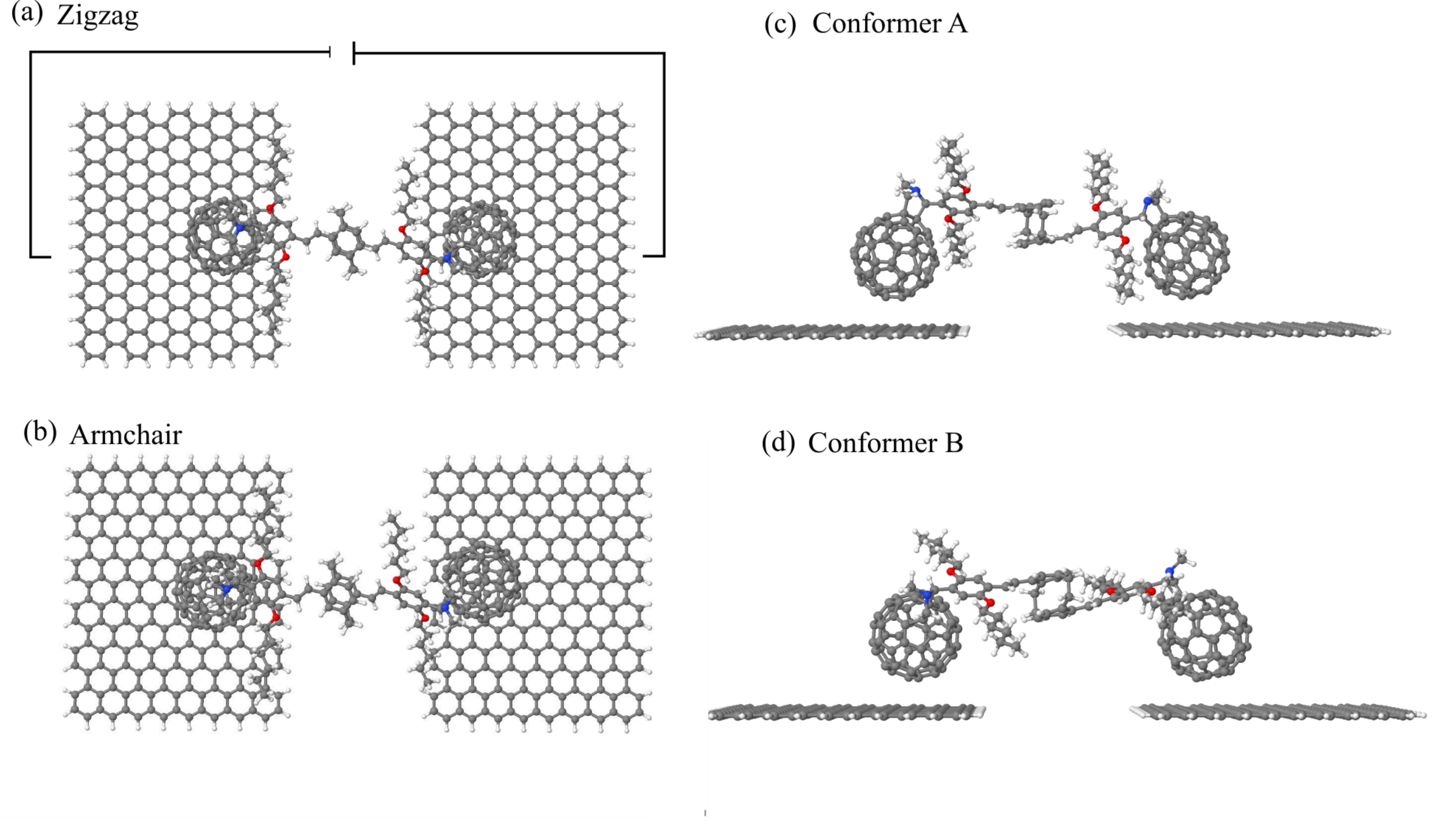}
    \caption{Single-molecule junctions investigated in this work. (a) Junction with zigzag-terminated graphene electrodes and (b) armchair-terminated graphene electrodes. 
      (c) Side view of molecular conformer A and (d) conformer B.}
\label{fig:systems}
\end{figure*}

\section{Systems and methods}  \label{sec:methods}

\subsection{Molecular junctions} \label{sec:methods_setup}

The architecture of the molecular junctions investigated is depicted in Fig.\ \ref{fig:001}. It comprises a C$_{\texttt{60}}$-endcapped molecular bridge with two styrene units that are covalently connected
via a [2,2']paracyclophane moiety in the center.\cite{Ontoria,Wielopolski} The molecular bridge binds via the fullerene anchor groups to two graphene leads. Based on this architecture, we have studied different types of molecular junctions depicted in Fig.\ \ref{fig:systems}.

For the transport calculations, extended systems are defined, comprising the molecule and a finite part of the graphene electrodes.
The graphene edges are saturated with hydrogen atoms in order to avoid dangling bonds.
The extended systems are coupled to two-dimensional semi-infinite graphene leads, described via absorbing boundary conditions (see Sec.\ \ref{sec:methods_transport}).
We investigate zigzag-terminated graphene nanocontacts as shown in  Fig.\ \ref{fig:systems} (a) as well as armchair termination,
Fig.\ \ref{fig:systems} (b). 
The C$_{\texttt{60}}$ anchor groups of the molecule are bounded to the graphene leads via dispersion interactions resulting in a distance of about 1.5 nm.

For both types of graphene edge termination, 
the geometry optimization (see Sec.\ \ref{sec:methods_elstruc}) reveals the existence of two structural conformations (A,B) corresponding to local minima of the potential energy surface, whose energies differ by 0.3 eV. The two conformations exhibit different orientations of the
phenylenevinylene moieties in the central [2,2']-paracyclophane-phenylenevinylene fragment (cf. Fig.\ \ref{fig:systems} c,d).

\subsection{Electronic structure methods} \label{sec:methods_elstruc}

The equilibrium geometries of the different single molecule junctions analyzed in this work were obtained using periodic density functional theory (DFT) calculations. Specifically, slab models 
consisting of two graphene leads and the molecular linker comprising 784 (834) atoms for the zigzag (armchair) terminated junctions were used. To prevent artificial 
interactions between successive slabs, a vacuum layer with a minimum width of 12 \AA \ was enforced. 
In all the optimizations, the molecular bridge and the hydrogens saturating the graphene edges were relaxed. 
The DFT calculations were carried out using the Perdew-Burke-Emzerhof (PBE) exchange-correlation 
functional,\cite{Perdew1996} including dispersion corrections employing Grimme's D3 method \cite{Grimme} and using the projected augmented wave method\cite{Blochl1994}. A 
k-mesh of 2$\times$3$\times$1 points was generated using the Monkhorst-Pack technique\cite{Monkhorst1976} and an 
energy cut-off of 
415.0 eV was used after testing. These calculations were carried out using VASP.\cite{VASP} 
Geometry relaxation was stopped when forces where smaller than 0.02 eV/A.
Following previous work\cite{Ivan2013}, transport calculations were carried out using cluster models built from the equilibrium structures obtained using the periodic DFT calculations. 
The sizes of the different clusters were selected to simulate as correct as possible the graphene leads while keeping the system tractable from a computational point of view. 
The electronic structure calculations for the cluster models, necessary to obtain the Hamiltonian for the different systems investigated, were performed using TURBOMOLE\cite{turbomole}
at the DFT level of theory employing the B3LYP functional \cite{becke:93} and the def-SV(P) basis set.\cite{schafer1992} 
Dispersion interactions were incorporated using Grimme's empirical dispersion correction.\cite{Grimme}

\subsection{Model and transport theory} \label{sec:methods_transport}

To study charge transport through the nanojunctions, we employ the calculation scheme presented in Ref.\ \citenum{Ivan2013}. 
This involves the partitioning of the Hamiltonian matrix describing the extended system into blocks corresponding to the molecular bridge, $H_m$,
the left and right contact, $H_l$ and $H_r$, and the respective couplings $V_{ml}, V_{mr},V_{lm}, V_{rm}$. \cite{Benesch08}
To model the effect of the semi-infinite graphene leads, we employ absorbing boundary conditions using complex absorbing potentials (CAPs) of the form $-iW_{\alpha}, \alpha=l,r$, 
with $(W_{\alpha})_{ij}=w_{\alpha i}\delta_{ij}$, which are modeled as
\begin{align}
 w_{\alpha i}=a|x_i-x_{\alpha}|^n
\end{align}
On the basis of test calculations, the prefactor $a$ was chosen such that the results do not depend on it. \cite{Saalfrank}
The absorbing potentials are added to the Hamiltonian of the graphene contacts, such that the overall Hamiltonian $\tilde H$ reads
\begin{equation}
\tilde H=
\begin{pmatrix}
H_{l}-iW_l & V_{lm} & V_{lr} \\
V_{ml} & H_m & V_{rm} \\
V_{rl} & V_{mr} & H_r -iW_r 
\end{pmatrix}.
\end{equation}
Thereby, the CAPs are applied in the graphene contacts far away from the fullerene molecules to avoid artificial effects.

Within the Landauer approach, transport through the contact is described by
the transmission function  $t(E)$, which for an electron with energy
$E$ is given by 
\cite{Datta}
\begin{equation}
 t(E)={\rm tr}\left\lbrace\Gamma_rG_m\Gamma_lG_m^{\dagger}\right\rbrace. 
\label{eqn:1}
\end{equation}
Hereby, $G_m$ denotes the Green's function projected onto the molecular bridge, 
\begin{equation}
 G_m(E)=\left[ E^+ - H_m-\Sigma_l(E)-\Sigma_r(E)\right]^{-1},
\end{equation}
which involves the self energies $\Sigma_{\alpha}(E)$
that describe the coupling of the molecular bridge to the left and 
right electrodes ($\alpha=l,r$), and are given by the expression
\begin{equation}
  \Sigma_{\alpha}(E)=V_{m\alpha}\left[E-H_{\alpha}+iW_{\alpha}\right]^{-1}V_{\alpha m}
\end{equation}
The self energies are related to the width-function 
$\Gamma_{\alpha}(E)$, used in Eq.\ (\ref{eqn:1}), via 
\begin{equation}
 \Sigma_{\alpha}(E)= \Delta_{\alpha}(E)-\frac{i}{2}\Gamma_{\alpha}(E),
\end{equation}
where $\Delta_{\alpha}$ denotes the level-shift function.
%In the wide-band limit, which is typically a good approximation 
%for gold electrodes, 
%$\Gamma_{\alpha}$ can be approximated by a constant value and 
%$\Delta_{\alpha}$ vanishes.

To characterize the molecular features in transport, we have in addition used a simpler approach, where the conductance properties are calculated for the isolated molecular bridge without explicit inclusion of the
graphene electrodes. 
In this approach, the electrodes are described within the wide-band approximation (WBA), 
where $\Delta_{\alpha}$ is zero and $\Gamma_{\alpha}$ is replaced by a constant. 
Specifically, the matrix elements of the self energies $\Sigma_{\alpha}(E)$ 
in a local atomic orbital basis are given by
\begin{equation}
 (\Sigma_{\alpha}(E))_{\nu\nu}=-\frac{i}{2}(\Gamma_{\alpha})_{\nu\nu}
 \label{wba}
\end{equation}
for orbitals $\nu$ corresponding to carbon atoms at the bottom of the C$_{\texttt{60}}$, 
which couple to the graphene surface. 
In the calculations reported below, we have used $(\Gamma_{\alpha})_{\nu\nu}= 1$ eV .

Based on the transmission function $t(E)$, the electrical current through
the junction is given by the expression
\begin{equation}
 I(V) = \frac{2e}{h}\int {\rm d}E \, t(E)(f_l(E)-f_r(E)),
\label{eqn:curr}
\end{equation}
where $f_{l/r}(E)$ denotes the Fermi function for the electrons in the
left/right lead.
It is given by
 \begin{equation}
 f_{\alpha}(E)= \frac{1}{1+e^{(E-\mu_{\alpha})/k_BT}}
 \end{equation}
 with the Boltzmann constant $k_B$, the electrode temperature $T$ and their chemical potentials $\mu_{\alpha}$ ($\alpha=l,r$). For a symmetric drop of the bias voltage $V$ around the Fermi energy $E_F$, the chemical potentials are given by
 \begin{equation}
  \mu_{\alpha}=E_F\pm \frac{eV}{2}.
 \end{equation}

\subsection{Analysis of transport pathways using local currents}  \label{sec:methods_loccurr}

Eqs.\ (\ref{eqn:1}) and (\ref{eqn:curr}) describe charge transport in terms
of the overall electrical current and transmission function of the complete junction.
In order to have a more detailed description and, in particular, 
to analyze pathways
of charge transport in the junction, we use  the method
 of local currents \cite{Todorov,solomon08}.
Within this technique
the overall current $I$ through a surface perpendicular to the 
transport direction
is represented in terms of local currents via
\begin{equation}
 I=\sum_{n'\in M_L}\sum_{n\in M_R} I_{n'n},
 \label{Iloc}
\end{equation} 
where $M_{L/R}$ denotes atomic sites left and right of the chosen 
surface, respectively. The local contributions 
to the current from site $n$ to $n'$ are given by\cite{solomon08,Pecchia04}
\begin{equation}
 I_{n'n}=\frac{2e}{h}\int {\rm d}E  K_{n'n}(E),
\label{eqn:localcurrs}
\end{equation}
with 
% \begin{equation}
%  K_{mn}(E)=\sum_{i\in m}\sum_{j\in n}(V_{ij}G_{ji}^<(E)-V_{ji}G_{ij}^<(E)).
% \label{eqn:2}
% \end{equation}
\begin{equation}
 K_{n'n}(E)=\sum_{\nu\in n'}\sum_{\mu\in n}
(V_{\nu\mu}G_{\mu\nu}^<(E)-V_{\mu\nu}G_{\nu\mu}^<(E)).
\label{eqn:2}
\end{equation}
Here, $V_{\nu\mu}$ denotes the coupling constant between orbitals $\nu$ and $\mu$, and $G^{<}$ is the lesser Green's function, given by
\begin{equation}
 G^{<}=(if_lG_m\Gamma_l G_m^{\dagger}
+if_rG_m\Gamma_r G_m^{\dagger}).
\end{equation}
For temperature $T=0$, the expression
$\sum_{n'\in M_L}\sum_{n\in M_R} K_{n'n}$ can be 
identified with the transmission function, Eq.\ (\ref{eqn:1}), 
and thus $K_{n'n}$ defines local contributions to the transmission
between pairs of atoms.

%\newpage
%\clearpage

\section{Results and discussion}\label{sec:results}

Using the methodology outlined above, we have characterized the transport properties of the molecular junctions. In the following, we analyze the transport properties based on the transmission function,  the relevant molecular orbitals and the local contributions to the transmission. A special emphasis is on the asymmetry of the transport characteristics with respect to bias polarity, observed in the experimental investigations, and the role of the different conformations as well as the type of termination of the graphene electrodes.

\subsection{Transport properties and molecular orbital analysis} \label{sec:results_mo}

\begin{figure}[t!]
	\centering
 	 \includegraphics[scale=0.75]{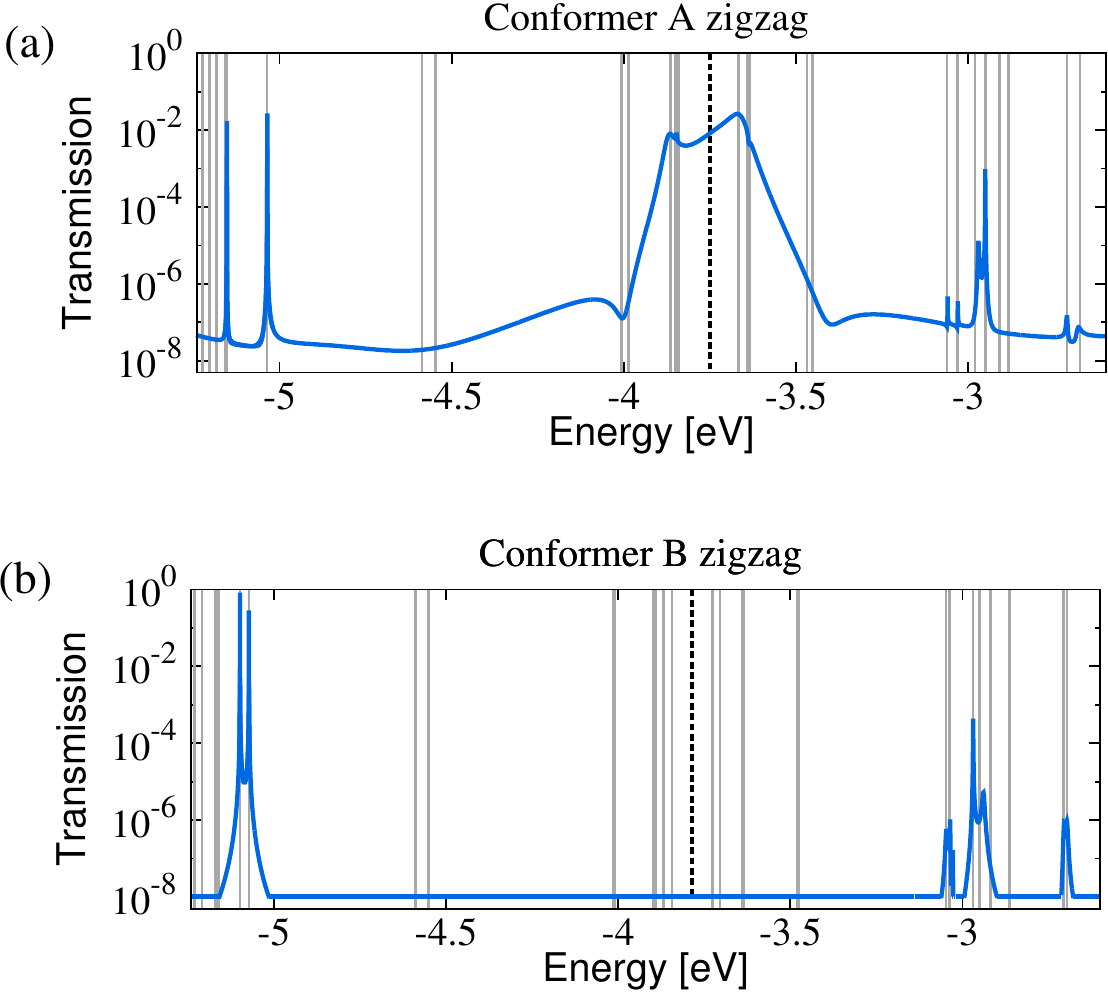}
\caption{Transmission function (blue line) for zigzag-terminated single-molecule junctions. The energy levels of the extended systems are depicted as grey lines. 
The black dotted line represents the Fermi level. (a) Conformer A and (b) conformer B.}
\label{fig:003}
\end{figure}

Fig.\ \ref{fig:003} shows the transmission function of the junctions with zigzag-terminated graphene electrodes for the two conformations of the molecular bridge (A and B), 
depicted in an energy range around the Fermi level, which in our simulation is identified with the center of the HOMO-LUMO gap. 
In addition, the energy levels are depicted as grey lines.
The transmissions exhibit a series of peak structures, which can be attributed to individual electronic levels of the extended junction as is outlined in the following.

\begin{figure}[h!]
\centering
\includegraphics[scale=0.3]{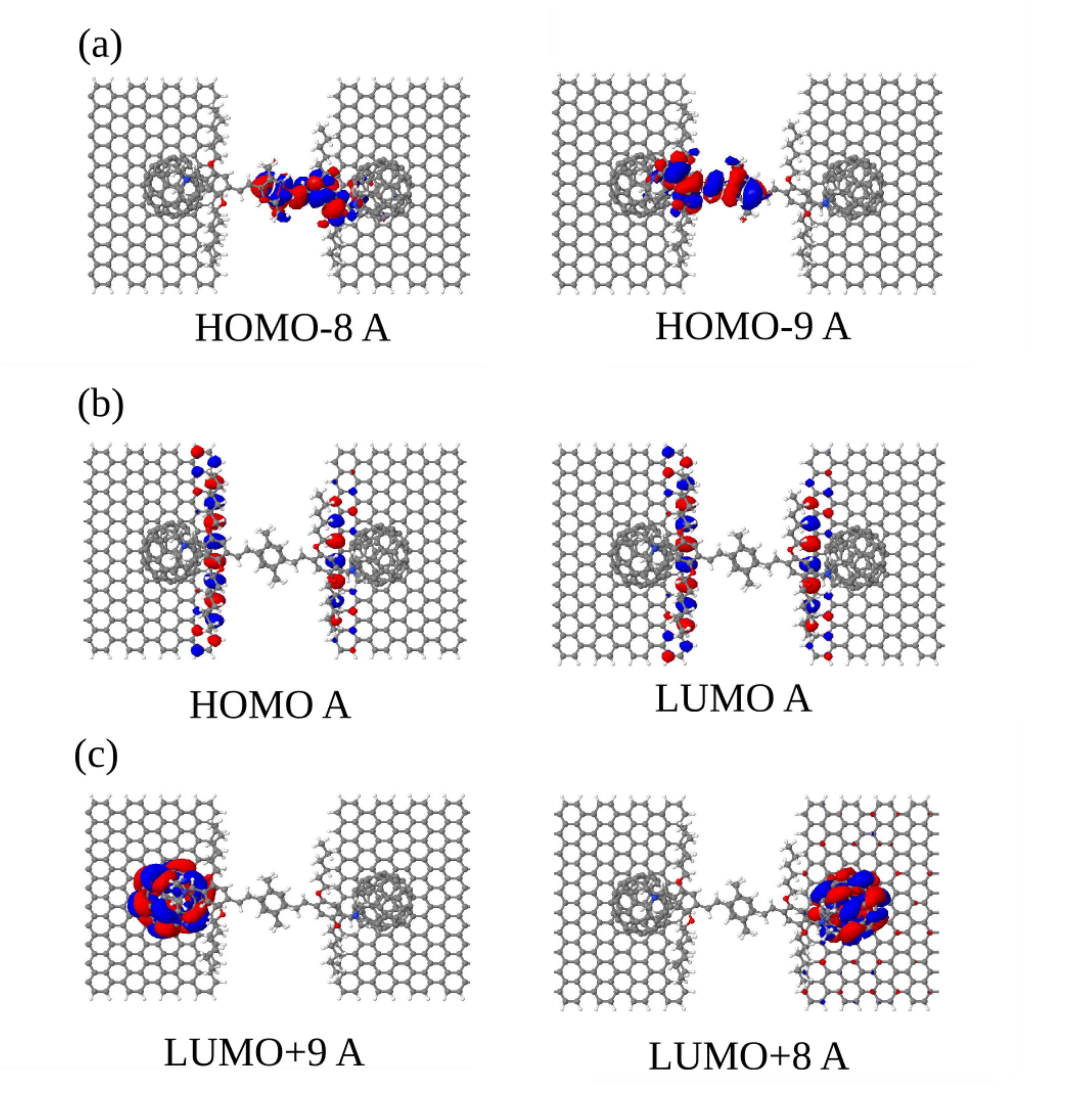}
\caption{Molecular orbitals of conformer A with zigzag terminated graphene electrodes.}
\label{fig:confAzigzag}
\end{figure}

The transmission function of conformer A (cf.\ Fig.\ \ref{fig:003} (a)) shows two narrow peaks below the Fermi level, 
at $E= -5.9$ and $E=-5.1$ eV, which are associated with the molecular levels HOMO-8 and HOMO-9 (of the extended molecular bridge) localized on the molecular bridge between the two fullerenes.
The corresponding molecular orbitals are shown in Fig.\ \ref{fig:confAzigzag} (a).
Due to the weak molecule-lead couplings, the resonance peaks are very narrow.
Close to the Fermi level, graphene edge states (cf. Fig.\ \ref{fig:confAzigzag} (b)) cause a broad peak structure with a maximum transmission value of $t=0.01$, which will be discussed in more detail in Sec.\ \ref{sec:interface}. 
Although this is a rather low transmission, these contributions would lead to a considerable current at low voltages, 
because the peak is significantly broader than those of the molecular resonances.
Above the Fermi energy at around $E=-3.0$ eV, a more complex resonance structure is found, which can be related to molecular states localized on the fullerene anchor groups, 
as shown in Fig.\ \ref{fig:confAzigzag} (e,f). The overall transmission value of these peaks is very low, the corresponding contribution to the current is thus negligible.

\begin{figure}
	\centering
	\includegraphics[scale=0.3]{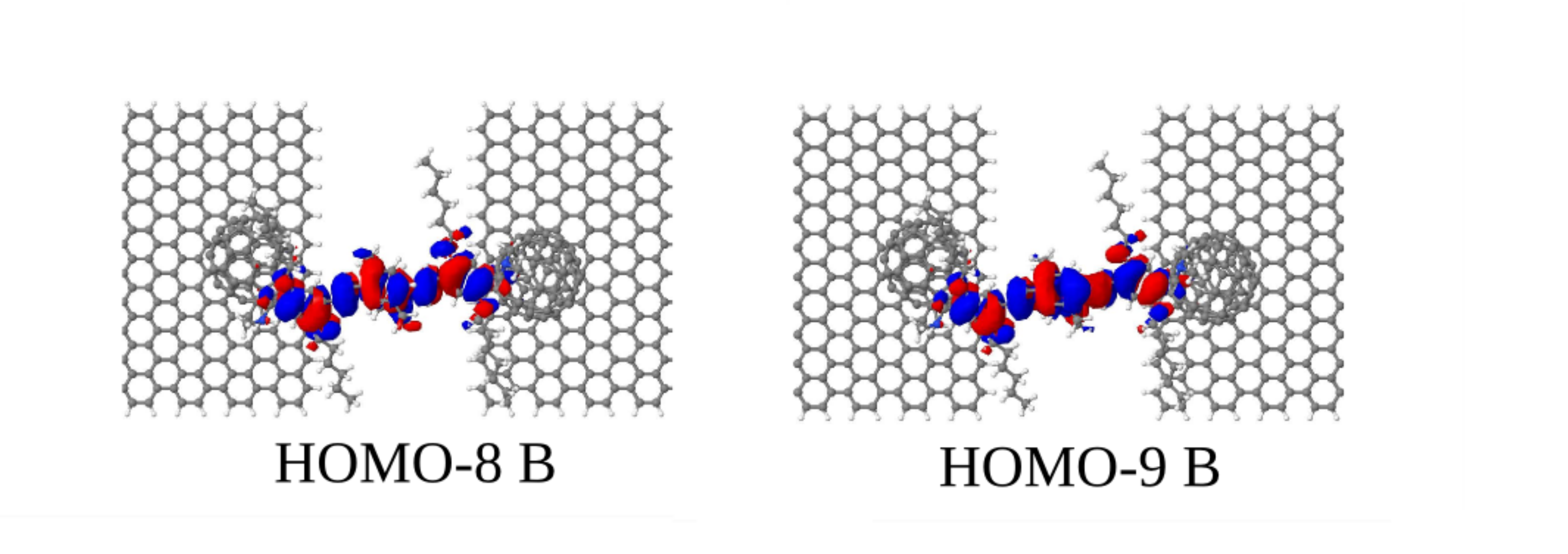}
  \caption{Molecular orbitals of conformer B with zigzag terminated graphene electrodes.}
\label{fig:confBzigzag}
\end{figure}

The transmission function of conformer B in Fig.\ \ref{fig:003} (b) also shows a double-peak structure at $-5.1$ eV due to states localized on the molecular bridge between the C$_{60}$ moieties. 
The corresponding molecular orbitals are depicted in Fig.\ \ref{fig:confBzigzag} (a,b).
The analysis of these orbitals for the two conformers reveals that they localize on opposite sides of the molecule, with those of conformer B being slightly more delocalized over the bridge.
The comparison of the transmission functions in Fig.\ \ref{fig:003} (a) and (b) demonstrates that the energy levels of these states are significantly closer for conformer B. 
The closer lying energy levels in combination with a more pronounced delocalization and thus higher transmission values result in an overall larger contribution of these double peak resonances to transport in conformer B compared to that in conformer A.
Despite having equivalent edge states at the Fermi level, no edge-induced transport contributions are present in the junction of conformer B.
As discussed in detail below, this is related to a difference in the molecule-lead coupling.
The transmission peak structure related to the molecular states on the fullerenes appears at similar energies as in conformer A, and its contribution to transport is also negligible.

\subsection{Role of asymmetry}

The experimental results presented in Ref.\ \onlinecite{Ullmann15} show a pronounced dependence of the current-voltage characteristics on the polarity of the bias voltage, i.e.\ asymmetric IV curves.
In Ref.\ \citenum{Ullmann15}, it could be established that the asymmetry is a result of contributions from the states localized on the molecular bridge between the two fullerenes. These states are HOMO-8 and HOMO-9 of the extended molecular bridge (Fig.\ \ref{fig:confAzigzag} (a) and Fig.\ \ref{fig:confBzigzag}), which correspond to HOMO and HOMO-1 of the isolated molecule. Thereby it was assumed (cf.\ SI of Ref.\ \citenum{Ullmann15}) that in the experimental setup used, the Fermi level in graphene is lowered to a similar value of that of gold (about -5 eV), 
such that the Fermi level lies in the proximity of the states.

Here, we extend the analysis of Ref.\ \citenum{Ullmann15} with respect to the asymmetry of the IV curves. To this end, we have investigated the response of the electronic levels relevant for transport (HOMO and HOMO-1 of the isolated molecule) to external electric fields applied along the line connecting the centers of mass of the fullerenes. 
A homogeneous electric field along this direction (cf.\ SI of Ref.\ \citenum{Ullmann15}) is included in the DFT calculations.
Thereby, as an estimate, the bias voltage was converted to an electric field using the fullerenes center of mass distance of 2.4 nm (electric field = bias voltage / 2.4 nm). Due to computational limitations, we have restricted the calculations to the isolated molecule without explicit inclusion of the graphene electrodes. The transmission functions for finite fields are obtained using the WBA for the coupling to the leads as described in Eq.\ (\ref{wba}).
In this way, the effect of the electric field in the junction for different bias voltage polarities is modeled and the polarizing effect of the applied bias can be included, which is important for the asymmetries in the transport properties.

\begin{figure}
    \centering
	  \includegraphics[scale=0.7]{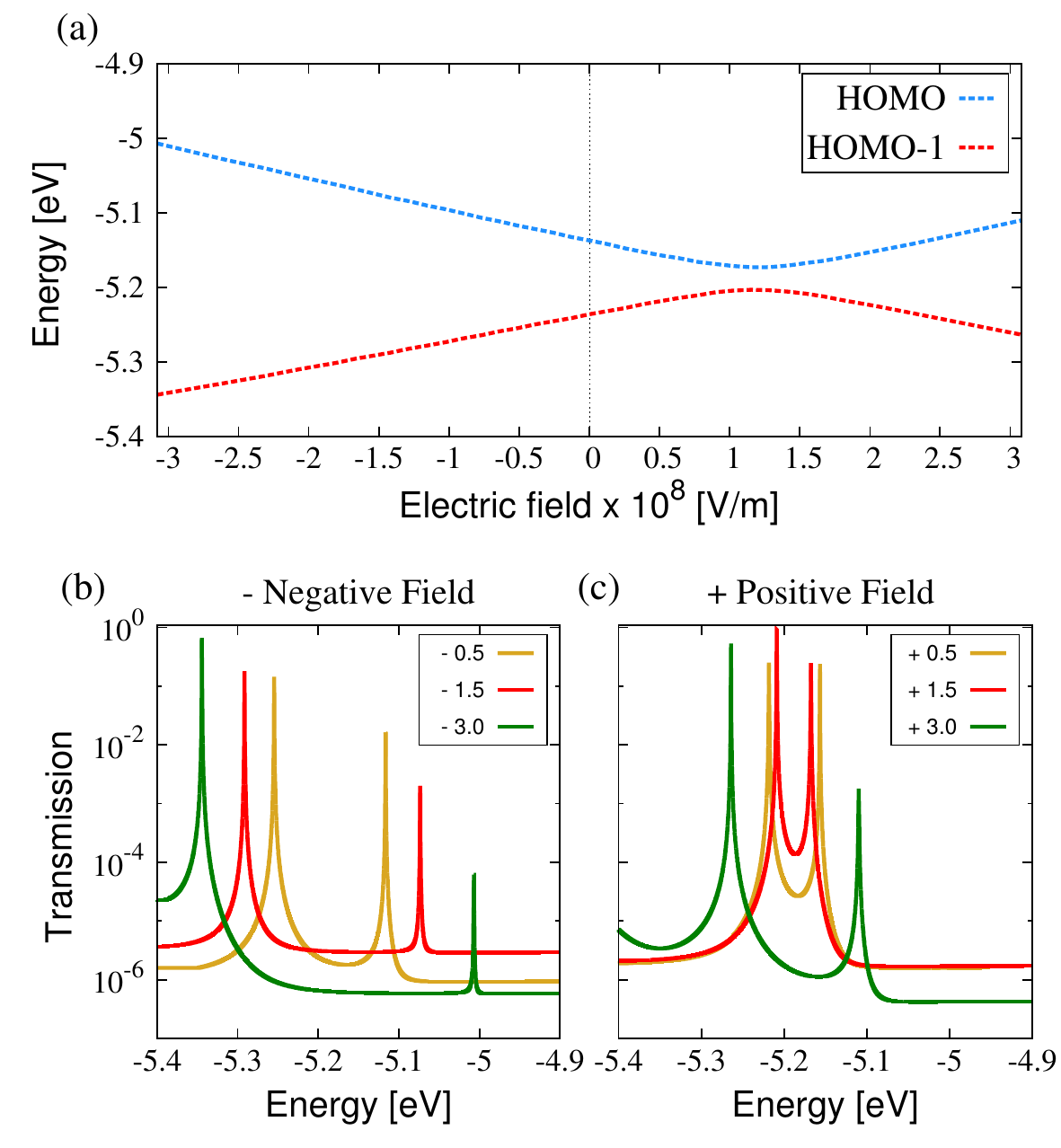}
	  \caption{(a) Field dependence of HOMO and HOMO-1 of the isolated molecular bridge in conformation A. (b) Field-dependence of the transmission function for a field of -0.5, -1.5 and -$3\cdot 10^{-8}$ V/m 
	  (c) and for positive fields.  }
	  \label{fig:trans-field}
\end{figure}

The analysis for conformer A is presented in Fig.\ \ref{fig:trans-field}. The dependence of the energy levels on the external electric field shows a typical avoided crossing with a minimum of the energy level spacing at a finite, positive field. As a consequence of the behavior of the energy levels, the double peak structure in the transmission function also depends sensitively on the electric field, as illustrated in Fig.\ \ref{fig:trans-field} (b) for negative and (c) for positive field. Specifically, for positive fields, the peaks between $-4.9$ to $-5.4$ eV are closer and the transmission is considerably enhanced, while for a negative field, the peaks are more separated and the transmission is significantly lower. Taking into account additional broadening due to coupling to vibrations, this is in agreement with the experimental results which shows for lower temperatures a double peak structure with a lower current for one bias polarity but a single peak structure with higher current for the other bias polarity.

As discussed in Ref.\ \citenum{Ullmann15}, the field dependence of the energy levels of conformer B is equivalent to that of conformer A for fields of equal strength but applied in opposite direction.
Therefore, a transition between the two conformations will revert the bias polarity dependence of the IV characteristics.
The inversions of the asymmetries in the IVs, observed in the experiment during consecutive voltage sweeps, can be assigned to conformational changes of the molecule,
namely a transition between conformer A and B.
It is interesting to note that for the molecular states on the fullerenes (cf.\ Fig.\ \ref{fig:confAzigzag} (c)), no distinct bias polarity dependence could be found.

\subsection{Junction with armchair termination}\label{sec:edge}

Graphene nanoribbons exhibit electronic states localized on the zigzag edge with energies located near the Fermi level.\cite{wakabayashi09}
In contrast, the armchair edge does not support such localized states. 
In molecule-graphene nanojunction, localized edge states can contribute to transport due to the interaction with adjacent molecular resonances.\cite{Ivan2013,Motta}
An example is the junction of conformer A with zigzag terminated graphene leads, which exhibits significant contributions due to edge states, cf.\ Fig.\ \ref{fig:confAzigzag}(b). 
%The interaction of the levels may be controlled by changing the %coupling between the leads and the molecule.\cite{Ivan2013}

\begin{figure}[t!]
	\centering
	\includegraphics[scale=0.7]{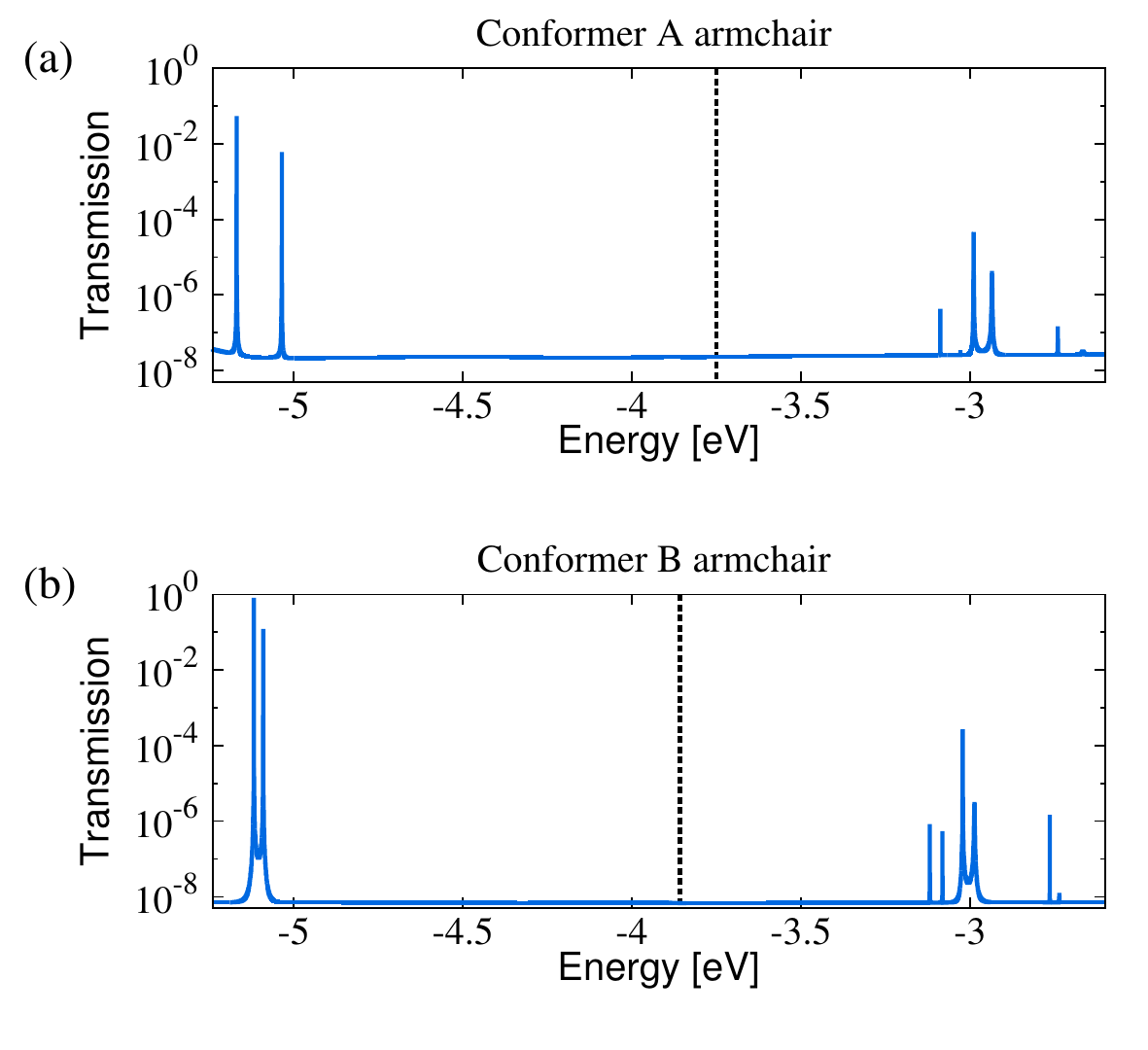}
\caption{Transmission function (blue line) for armchair terminated single-molecule junctions. 
The black dotted line represents the Fermi level. (a) Conformer A and (b) conformer B. }
\label{fig:transarm}
\end{figure}

Here, we analyze for comparison the transport properties of graphene-molecule junctions with armchair termination of the graphene leads, as shown in Fig.\ \ref{fig:systems} (c). The geometry optimization again yields two conformations A and B with an energy difference of $\sim 0.3$ eV.
The transmission functions for the two conformers with armchair termination of the electrodes are presented in Fig.\ \ref{fig:transarm}.
The results show that the contributions of the molecular states related to  HOMO and HOMO-1 (at about $-5.1$ eV) of the isolated molecule and those induced by the states localized on the fullerenes (at about $-3.0$ eV) are almost independent on the type of graphene termination. However, the contributions from edge states close to the Fermi energy, which are present in the zigzag-terminated junction of conformer A, are missing in the armchair terminated junction for both conformers. 
In the experiment of Ref. \citenum{Ullmann15}, the existence of a well-defined zigzag termination is rather unlikely and thus edge-state transport channels are expected to be less relevant.

\subsection{Interface geometry}\label{sec:interface}

The fact that edge-state induced transport contributions are found in the junction in conformer A with zigzag termination, but not in conformer B, can be related to a difference in the interface geometry and the resulting coupling between the molecule and the graphene leads.
These investigations build on previous experimental and theoretical studies for related systems, which revealed significantly different adsorption energies, depending on the orientations, of C$_{\texttt{60}}$ on graphene. \cite{kimlee}

\begin{figure}[t!]
%\begin{minipage}[h!]{7.5cm}
	\centering
	\includegraphics[scale=0.25]{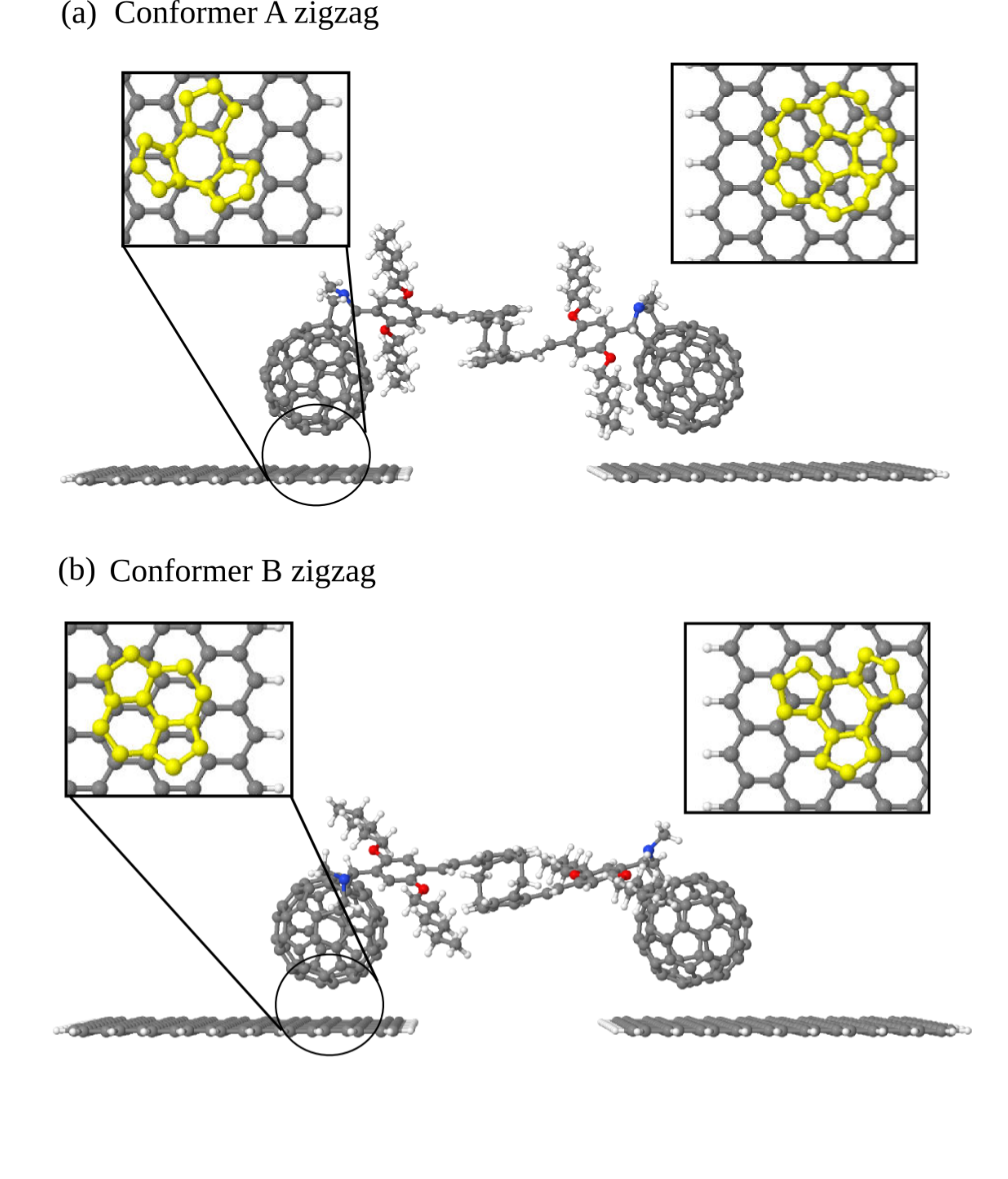}
%\end{minipage}
\caption{Interface geometry of (a) conformer A and (b) conformer B. In (a), the interface on the left hand side corresponds to a hexagon/hole geometry, while on the right hand side it resembles a pentagon/hole geometry. 
In (b), the left hand side interface corresponds to a dimer/bridge geometry,
and at the right hand side to a hexagon/hole geometry. }
\label{fig:008}
\end{figure}

Fig.\ \ref{fig:008} shows the interface geometry for the two conformers. 
In the junction of conformer A, the molecule binds on the left hand side via a hexagon directly oriented to the center of a graphene hexagon, 
which, using the notation of Ref.\ \citenum{kimlee}, corresponds to a hexagon/hole geometry. 
The interface on the right hand side closely resembles a pentagon/hole geometry (cf.\ Fig.\ \ref{fig:008} (a)). 
For conformer B (Fig.\ \ref{fig:008} (a)) the left hand side interface corresponds to a dimer/bridge geometry,
whereas at the right hand side it is a hexagon/hole geometry. 
While for the hexagon/hole and pentagon/hole orientation stronger binding energies were found, the dimer/bridge geometry was observed to be less stable.
In accordance with Ref.\ \citenum{kimlee}, this will lead to a lower coupling in conformer B, preventing the hybridization of edge states to become conducting states.
Moreover, as a consequence of the different orientations of the fullerenes relative to the graphene surface, also the coupling to the edges differs.
On the right hand side in conformer A, a pentagon is oriented towards the graphene surface, which is surrounded by hexagons pointing towards the edge with zigzag termination. 
On the other hand, in conformer B a hexagon is pointing to the surface,  whereas a hexagon and a pentagon couple to the zigzag edge. In the latter case, a weaker interaction is expected.

\begin{figure*}[t!]
\centering
\includegraphics[scale=0.99]{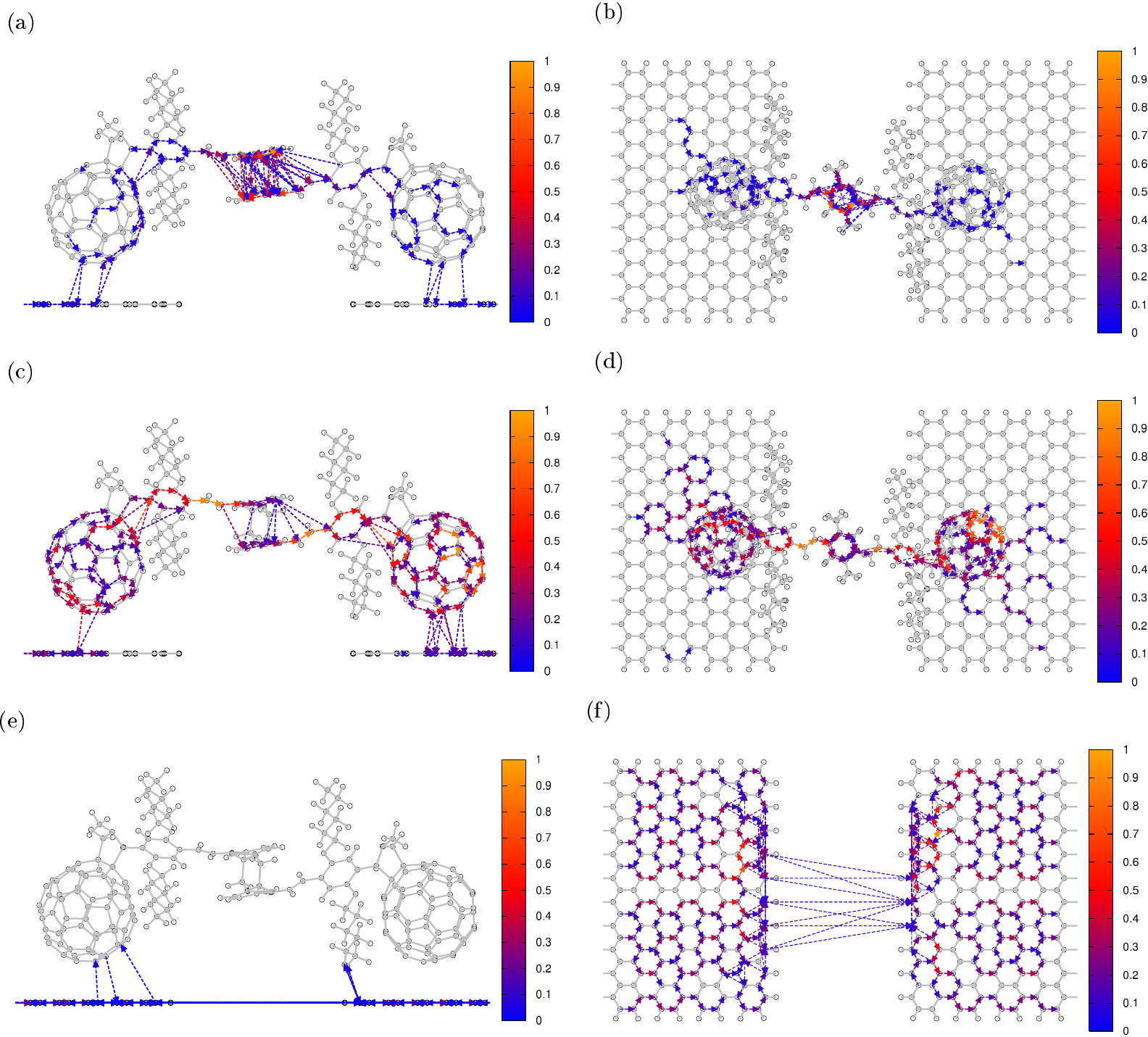}
\caption{Local transmissions (Top and side-view) at (a,b) $E_{\text{HOMO-8}}$, (c,d) $E_{\text{LUMO+8}}$, (e,f) $E_{\text{LUMO}}$ (top view plotted without molecule). }
\label{fig:localtrans-noncov}
\end{figure*}

\subsection{Local transmission paths}

Finally, we analyze the pathways of electrons in the molecule-graphene junctions. To this end, the local transmissions as defined in Eq.\ (\ref{eqn:2}) are calculated at energies, where the total transmission is maximal. The results for conformer A and zigzag-terminated graphene leads are shown in  Fig.\ \ref{fig:localtrans-noncov} for a selection of representative energies.
Thereby, local transmissions between all atoms are depicted as arrows. The color code ranging from blue
to orange indicates the local transmission value. 
An arrow is only drawn when this value exceeds 10$\%$ of the maximal local transmission value, which for clarity is normalized for each energy.
Although the sum of all local transmissions across a line vertical to the transport direction is preserved in the calculations,
this conservation is not fully apparent in the plots due to the 10$\%$ of the transmission contributions which are not shown.
It is also noted that the transmission probability is absorbed in the leads beyond a certain region due to the CAP applied.

We first consider the local transmissions at $E_{\text{HOMO-8}}$, depicted in Fig.\ \ref{fig:localtrans-noncov} (a,b), as an example for transport through the occupied molecular levels.
In this case, the transmission pathway leads from the left electrode into the C$_{\texttt{60}}$ anchor group, from where it goes through the molecular wire into the other C$_{\texttt{60}}$ and finally reaches the right graphene electrode.
On the left hand side, the pathway enter the molecule via a C$_{\texttt{60}}$ hexagon, and on the right hand side via a pentagon, in correspondence with the C$_{\texttt{60}}$-graphene interface shown in Fig.\ \ref{fig:008} (a).
Considering the corresponding molecular orbital (cf.\ Fig.\ \ref{fig:confAzigzag}(a)), the non-vanishing local transmission components are mainly between atoms which exhibit significant electron density. The pathway at energy $E_{\text{HOMO-9}}$ follows a similar pattern (not shown).

Fig.\ \ref{fig:localtrans-noncov} (c,d) shows the local transmissions through the unoccupied molecular resonances, in particular $E_{\text{LUMO+8}}$. 
Since the electron density of LUMO+8 is concentrated on the fullerenes (cf.\ Fig.\ \ref{fig:confAzigzag}(c)), the local transmissions are spread over the C$_{\texttt{60}}$ moieties.

A quite different pattern is obtained for the local transmission evaluated at the energy of the edge-induced peak around the Fermi level, depicted in  Fig.\ \ref{fig:localtrans-noncov} (e,f). In this energy region, the local transmission exhibits a complex pattern involving pronounced circular contributions in the graphene contacts, transmissions which are back scattered at the edges as well as direct transitions from one lead to the other.
Transmissions into the molecule are vanishingly small, as shown in the side view of the junction in Fig.\ \ref{fig:localtrans-noncov} (e) and involve contributions between the graphene and the side chain of the molecule. Small contributions along the molecule are not seen because their value is lower than the cutoff of 10$\%$.

While local ring-shaped current structures are a well known phenomenon in graphene nanoribbons \cite{walz2014,wilhelm2014}, direct transitions between graphene leads have not been analyzed so far. Test calculations, which artificially neglect the molecular states, do not show these direct transitions, thus demonstrating that they are induced by the hybridization between the edge states and the molecular states. This finding can be further confirmed by a qualitative consideration of the potential barrier.
Without molecule, the potential barrier for an electron to tunnel through the junction with a vacuum gap of 1.5 nm is too high to give noticeable transmission values.
With molecule in conformation A attached, however, the energy barrier is decreased significantly thus facilitating direct tunneling transitions.

\section{Conclusions}

We have investigated charge transport 
in C$_{\texttt{60}}$-based single-molecule junctions with graphene electrodes. These type of molecule-graphene junctions were recently studied experimentally using epitaxial single layer graphene.\cite{Ullmann15}
Our theoretical study was based on a combination of DFT electronic structure calculations and 
Landauer transport theory. Furthermore, the theory of local currents was used to analyze pathways of electrons through the junctions.
The results of electronic structure calculations show that the molecular bridge can exist in two different conformations, which are relevant for charge transport. In addition, we have considered molecular junctions with zigzag and armchair termination of the graphene electrodes.

The analysis of the transport characteristics in terms of molecular orbitals and local transmissions reveals the existence of two different types of transport channels. The first type includes transport through molecular orbitals, which are either localized on the molecular bridge or on the C$_{\texttt{60}}$ anchor groups. Thereby, the latter contribution is of less relevance in the junctions investigated. In junctions with zigzag-termination, furthermore, edge-state induced transport channels contribute. 
The results also show that, for the molecule-based transport channels, the two different conformers exhibit a reversed dependence of the transport characteristics on the bias polarity, which was also found in experiment. Furthermore, the different interface geometry of the two conformers causes a significant difference in molecule-lead coupling. As a consequence, edge-induced transport channels are only present in one of the conformers.

As mentioned in the introduction, a major advantage of graphene as material for electrodes in molecular junctions if the possibility to access the junction with light to monitor and control the nonequilibrium state of the junction. The theoretical description of this scenario is an equally interesting and challenging extension of the present work.

\section*{Acknowledgments}
This work has been supported by the
the Deutsche Forschungsgemeinschaft (DFG) through the Cluster
of Excellence 'Engineering of Advanced Materials' (EAM) and
SFB 953. Generous allocation of
computing time at the computing centers in Erlangen (RRZE),
Munich (LRZ), and J\"ulich (JSC) is gratefully acknowledged.
The molecules were provided by Augustin Molina-Ontoria and Nazario Martin.

\end{document}